\documentstyle[amssymb,pra,aps,manuscript]{revtex}
\input{epsf}

\begin{document}
\draft
\title{Multiparticle Quantum Superposition \\
and Stimulated Entanglement\\
by Parity Selective Amplification of Entangled States\\
}
\author{Francesco De Martini and Giovanni Di Giuseppe}
\address{Dipartimento di Fisica and Istituto Nazionale di Fisica della Materia \\
Universit\'{a} ''La Sapienza'', Roma 00185, Italy}
\maketitle

\begin{abstract}
A multiparticle{\it \ }quantum superposition state has been generated by a
novel phase-selective parametric amplifier of an entangled two-photon state.
This realization is expected to open a new field of investigations on the
persistence of the validity of the standard quantum theory for systems of
increasing complexity, in a quasi {\it decoherence-free} environment.
Because of its nonlocal structure the new system is expected to play a
relevant role in the modern endeavor on quantum information and in the basic
physics of entanglement.
\end{abstract}

\pacs{PACS numbers: \tt\string03.65.Bz, 03.67.-a, 42.50.Ar, 89.70.+c}

\narrowtext

Since the golden years of quantum mechanics the interference of classically
distinguishable quantum states, first introduced by the famous
''Schroedinger Cat'' apologue \cite{1}, has been the object of extensive
theoretical studies and recognized as a major conceptual paradigm of physics 
\cite{2,3}. In modern times the sciences of quantum information and quantum
computation deal precisely with collective processes involving a
multiplicity of interfering states, generally mutually ''entangled'' and
rapidly de-phased by decoherence \cite{4}. For many respects the
experimental implementation of this intriguing classical-quantum condition
represents today an open problem in spite of recent successful studies
carried out mostly with atoms \cite{5,6,7}. A nearly decoherence-free
all-optical scheme based on the process of the {\it quantum injected}
optical parametric amplification (QIOPA) of a {\it single} photon in a
quantum superposition state, i.e, a {\it qubit}, has been proposed \cite{8,9}%
. As a relevant step forward in the realization of the quantum injection
scheme, the present work reports a novel optical parametric amplifier (OPA)
system that transforms any input linear-polarization ($\pi $) {\it entangled,%
} {\it 2-photon state} ({\it ebit}){\it \ }into a quantum superposition of $%
\pi $-{\it entangled,} {\it multi-photon states}, indeed an optical ''{\it %
Schroedinger Cat''} state (S-Cat) \cite{10}. In order to achieve this result
the new system implements an efficient parity-selective device, usually
referred to as ''nonlocal entangled interferometer'' (NEIF) \cite{11}. In
the language of electrical engeneering, NEIF\ conveys on different output
channels the {\it squeezed-vacuum} ''{\it noise''} and the ''{\it signal'', }%
viz. the amplified ebit state. This results in the generation of a S-Cat
state with a signal-to-noise ratio (S/N) which may be large, virtually
infinite.

Consider the experimental arrangement shown in Figure 1. A nonlinear (NL)
beta barium borate (BBO)\ crystal slab with parallel anti-reflection coated
faces, cut for Type II phase-matching and 1.5 mm thick, was excited in both
''{\it left}'' (L-) and ''{\it right}'' (R-) directions by an UV mode-locked
laser beam which was back-reflected by a spherical UV coated rear mirror $%
M_{UV}$ with curvature radius ($cr_{p}$) = 30 cm. Precisely, the L- (or R-)
amplification is the one determined by the UV beam directed towards the {\it %
left} (or {\it right}) in the Fig.1. %\begin{figure}[tbp]
%\epsfxsize=3.4in \epsfysize=2.6in
%\epsffile{PRL-ParitySelec_Fig_1.eps} \vglue 0.2cm
%\caption{Experimental apparatus.}
%\label{PRL-ParitySelec_Fig_1.eps}
%\end{figure}
A computer controlled mount allowed micrometric displacements of $M_{UV}$
along the axis {\bf Z, }parallel to the wavevector (wv) of the UV beam, {\bf %
k}$_{p}$. The UV beam was created by second-harmonic- generation of the
output of a Ti:Sa Coherent MIRA laser emitting pulses at the wavelength (wl) 
$\lambda _{p}=397.5nm$ with a coherence time $\delta _{t}=180fs$ at a 76 $%
{\it Mhz}$ rep-rate and with an average power of 0.3 W.

Consider first the L-amplification of the input vacuum state, viz. the
spontaneous parametric down conversion (SPDC)\ process. This was excited by
focusing the UV beam on the right plane surface of the NL crystal by a lens
with focal length $f_{p}=1m$. Photon couples in $\pi $-entangled states at a
wl $\lambda =795nm$ were then generated with an entanglement phase $\Phi
^{\prime }$ equal to\ the {\it intrinsic} QIOPA\ phase $\Psi $ determined by
the spatial orientation of the Type II crystal: $\Phi ^{\prime }=\Psi $ \cite
{11,12}. The dynamical role{\it \ }of $\Psi $ will be defined by the theory
given below. The state of each L-emitted photon couple could be then
generally expressed as $\left| \Phi ^{\prime }\right\rangle =2^{-\frac{1}{2}%
}[\left| 11.00\right\rangle +\exp (i\Phi ^{\prime })\left|
00.11\right\rangle ]$ where the state $|n_{1}n_{2}.n_{3}n_{4}\rangle \equiv $
$|n_{1h}\rangle |n_{2v}\rangle |n_{1v}\rangle |n_{2h}\rangle $ expresses the
particle occupancies of the Fock states associated with the relevant {\bf k}%
-modes, ${\bf k}_{j}$ () with horizontal $(h)$ or vertical $(v)$ linear
polarizations $(\pi )$. It is well known that the two optical modes
belonging to each couple $\left\{ 1h,2v\right\} $ and $\left\{ 1v,2h\right\} 
$ are parametrically correlated, either for L- and R-amplifications,
respectively by two {\it equal} and {\it mutually} {\it independent}
amplifiers $OPA_{A}$ and $OPA_{B}$, the ones that implement the overall
action of any Type II $(OPA)$ operating in non-degenerate mode configuration 
\cite{8,9,11}. In the experiment each L-emitted photon pair was selected by
a couple of pinholes, back-reflected and re-focused onto the left surface of
the NL crystal by two equal spherical mirrors $M_{j}$ ($j=1,2$) with
reflectivity 100\% at $\lambda $ and radius $cr_{\lambda }$ =30 cm. Both $%
M_{j}$ were placed at an adjustable distance l $\simeq $ 30 cm from the
source crystal. Care was taken to precisely overlap in the NL\ slab the
focal regions (with diameter $\phi \approx 70\mu m$) of the two back
reflected beams at $\lambda $ and of the back reflected UV beam at $\lambda
_{p}$. During the two photon back-reflection and before {\it re-injection}
of the couple into the QIOPA, a ${\it \lambda /4}$ plate with angular
orientation $(\varphi )$ introduced a change of the entanglement phase of $%
\left| \Phi ^{\prime }\right\rangle $: $\Phi ^{\prime }\rightarrow \Phi $
i.e. the original state of the photon couple, $\left| \Phi ^{\prime
}\right\rangle $ was transformed just before re-injection into:

\begin{equation}
\left| \Phi \right\rangle =2^{-\frac{1}{2}}[\left| 11.00\right\rangle +\exp
(i\Phi )\left| 00.11\right\rangle ]\equiv 2^{-\frac{1}{2}}[\left| \uparrow
\right\rangle _{1}\left| \downarrow \right\rangle _{2}+\exp (i\Phi )\left|
\downarrow \right\rangle _{1}\left| \uparrow \right\rangle _{2}]
\end{equation}

The labels $j=1,2$ of the orthogonal states in the above spin $(\pi )$
representation refer to the optical modes ${\bf k}_{j}$. Let's assume now
the precise simple conditions adopted in the experiment: $\varphi =0$ and: $%
\Psi =\Phi ^{\prime }=0$. As the phase shifting action of the ${\it \lambda
/4}$ plate for $\varphi =0$ leads to $\Phi \rightarrow \Phi ^{\prime }+\pi $
and then, in our case to $\Phi =\pi $, the originally L-generated {\it %
triplet} $\left| \Phi ^{\prime }\right\rangle $ with $\Phi ^{\prime }=0$ was
re-injected into the R-amplifier phase-transformed into the {\it odd-parity} 
{\it singlet}: $|\Phi \rangle _{s}$=$2^{-{\frac{1}{2}}}[|11.00\rangle
-|00.11\rangle ]$. As we shall see shortly, this {\it odd} {\it parity }ebit 
$|\Phi \rangle _{s}$ is finally R-amplified by QIOPA\ into a quantum
superposition of {\it odd-parity, multi-photon pure} S-Cat states: $\left|
\Phi \right\rangle _{OUT}$=$2^{-{\frac{1}{2}}}[|\Psi _{A}\rangle -|\Psi
_{B}\rangle ]$.

In order to complete the overall argument and to clarify the physical origin
of the (S/N) selectivity of the system, consider now the R-amplification of
the {\it input vacuum }state $|vac\rangle \equiv |00.00\rangle $ i.e. the
transformation of $|vac\rangle $ into the output ''{\it squeezed vacuum''}
state \cite{8,9}. A quantum analysis shows that this state is represented by
a {\it thermal distribution} where the {\it parity} of the entangled Fock
states appearing in the sum is, once again, determined by the {\it intrinsic}
phase $\Psi $ \cite{13}. Since in our experiment we set $\Psi =0$, the
squeezed vacuum state finally consists of a sum of {\it even-parity} states: 
$|\Psi _{vac}\rangle _{OUT}\simeq -C^{-2}[|00.00\rangle $ + $\Gamma
(|11.00\rangle +|00.11\rangle )$ + $\Gamma ^{2}|11.11\rangle $+ $\Gamma
^{2}(|22.00\rangle +|00.22\rangle )$+...$]$ where: $C\equiv \cosh g\approx 1$%
, $\Gamma =\tanh g<1$ and the parametric\ ''gain'' $g$ are dynamical
parameters adopted in the $OPA$ analysis below.

In summary, in our experiment the output ''{\it signal''}, i.e. the {\it R}-%
{\it amplified} {\it ebit}, consists of a superposition of {\it odd- parity}
states while the output ''{\it noise}'', i.e. the {\it squeezed vacuum,}
consists of a superposition of {\it even-parity} states. At last NEIF
provides the selective addressing to different output channels of the
entangled states having different symmetries, i.e. here the ones expressing
respectively the {\it signal} and the {\it noise}. This is the key idea
underlying the parity-sensitive, post-selective properties of the system 
\cite{8}.

Let us now analyze in more details the QIOPA, viz. the R-amplification
process \cite{13}. The two independent amplifiers $OPA_{A}$ and $OPA_{B}$
implementig the overall $OPA$\ process induce unitary transformations
respectively on two couples of\ time $(t)$ dependent field operators: $\hat{a%
}_{1}(t)\equiv \hat{a}(t)_{1h}$, $\hat{a}_{2}(t)\equiv \hat{a}(t)_{2v}$ and $%
\hat{b}_{1}(t)\equiv \hat{a}(t)_{1v}$, $\hat{b}_{2}(t)\equiv \hat{a}(t)_{2h}$
for which, at the initial interaction $t$ and for any $i$ and $j$ and $%
i,j=1,2$ is: $[\hat{a}_{i},\hat{a}_{j}^{\dagger }]=[\hat{b}_{i},\hat{b}%
_{j}^{\dagger }]=\delta _{ij}$ and $[\hat{a}_{i},\hat{b}_{j}^{\dagger }]$ $%
=0 $, being: $\hat{a}_{i}\equiv \hat{a}_{i}(0)$, $\hat{b}_{i}\equiv \hat{b}%
_{i}(0)$ the field operators at the initial interaction time $t=0${\it .}
The Hamiltonian of the interaction is expressed in the form: $H_{I}=i\hbar
\chi \lbrack \hat{A^{\dagger }}+e^{i\Psi }\hat{B^{\dagger }}]+h.c.$ where: $%
\hat{A^{\dagger }}\equiv \hat{a}_{1}(t)^{\dagger }\hat{a}_{2}(t)^{\dagger }$,%
$\ \hat{B^{\dagger }}\equiv \hat{b}_{1}(t)^{\dagger }\hat{b}_{2}(t)^{\dagger
}$, $g$ $\equiv \chi t$ is a real number expressing the {\it amplification} 
{\it gain, }and{\it \ }$\chi $ the coupling term proportional to the product
of the 2$^{nd}$-order NL susceptibility of the crystal and of the {\it pump}
field, here assumed ''classical'' and undepleted by the interaction. The
interaction $t$ may be determined in our case by the length l of the NL
crystal. The quantum dynamics of \ $OPA_{A}$ and $OPA_{B}\ $is expressed by
the mutually commuting, unitary {\it squeeze operators}: $U_{A}(t)=$ $exp[g(%
\hat{A^{\dagger }}-\hat{A})]$ and $U_{B}(t)=$ $exp[ge^{i\Psi }(\hat{%
B^{\dagger }}-\hat{B})]$ implying the following Bogoliubov transformations
for the field operators: $\hat{a}_{i}(t)=C\hat{a}_{i}+S\hat{a}%
_{j}{}^{\dagger }$; $\hat{b}_{i}(t)=C\hat{b}_{i}+\widetilde{S}\hat{b}%
_{j}{}^{\dagger }$ with $i\neq j$ \cite{8,9}. Here: $S$ $\equiv \sinh g$, $%
\widetilde{S}\equiv $ $e^{i\Psi }S$.

Of course the same dynamics holds for the L-amplification, viz. the SPDC
process, generally with a different value of the gain: $g=\eta g^{\prime }$, 
$\Gamma \simeq \eta \Gamma ^{\prime }$ being the scaling parameter: $\eta
\simeq (f_{p}/r_{p}$) and assuming that primed and umprimed parameters refer
to the processes of L- and R-amplifications, respectively. The adoption of a
scaling parameter $\eta >1$ was found to represent a relevant experimental
resource as a larger $\eta $ leads comparatively to: (a) A larger gain $g$
of the QIOPA, R-amplification: leading to a larger gain effect. (b) A
smaller gain $g^{\prime }$ of the SPDC,\ L-amplification: implying a smaller
emission rate of unwanted SPDC {\it double} photon couples. With the adopted
value $\eta =3$ the ratio of the SPDC rate of unwanted {\it double} photon
couples was $10^{-2}$ smaller than the rate of {\it single} couples, the
ones that after back-reflection ad phase-transformation are expressed by the
input state $|\Phi \rangle _{s}$.

Let us return to the R-amplification process. By the use of the evolution
operator $U_{AB}(t)$=$U_{A}(t)U_{B}(t)$ and of the {\it disentangling theorem%
} the quantum injection of the input state given by Eq. 2 leads to a {\it %
Schroedinger-Cat} form for the output state:

\begin{equation}
\left| \Phi \right\rangle _{OUT}=U_{AB}(t)\left| \Phi \right\rangle =2^{-{%
\frac{1}{2}}}[|\Psi _{A}\rangle +e^{i\Phi }|\Psi _{B}\rangle ]
\end{equation}
which, in agreement with the original definition \cite{1,2}, is expressed
here as the quantum superposition of the following {\it multi-particle }%
states: 
\[
|\Psi _{A}\rangle \simeq \sqrt{2}\eta C^{-5}\sum_{n,m:0}^{\infty }n\Gamma
^{(n+m)}\left| nn.mm\right\rangle ;\ |\Psi _{B}\rangle \simeq \sqrt{2}\eta
C^{-5}\sum_{n,m:0}^{\infty }m\Gamma ^{(n+m)}\left| nn.mm\right\rangle 
\]

Note that the phase $\Phi $ of the input state, Eq. 1, and then its parity
is reproduced into the {\it output} multiparticle state and determines the
quantum superposition character of the S-Cat. This {\it phase preserving}
property appears to be a common feature of all parametric
amplification/squeezing transformations of entangled quantum states \cite
{8,9}.

We may also inspect the superposition status of the S-Cat by investigating
the Wigner function of $\left| \Phi \right\rangle _{OUT}$. We first evaluate
the {\it symmetrically }ordered {\it characteristic function} of the set of
complex variables $(\eta ,\eta ^{\ast },\xi ,\xi ^{\ast })\equiv \{\eta ,\xi
\}$: $\chi _{_{S}}\{\eta ,\xi \}$ = $\langle \Phi |D[\eta (t)]D[\xi
(t)]|\Phi \rangle $ expressed in terms of the {\it displacement }operators $%
D[\eta (t)]\equiv \exp [\eta (t)\hat{a}(0)^{\dagger }-\eta ^{\ast }(t)\hat{a}%
(0)]$ and${\it \ }D[\xi (t)]\equiv \exp [\xi (t)\widehat{b}(0)^{\dagger
}-\xi ^{\ast }(t)\widehat{b}(0)]$ where: $\eta (t)\equiv (\eta C-\eta ^{\ast
}S)$; $\xi (t)\equiv (\xi C-\xi ^{\ast }S)$\cite{8}. The Wigner function $%
W\{\alpha ,\beta \}$ of the complex {\it phase-space variables} $(\alpha
,\alpha ^{\ast },\beta ,\beta ^{\ast })\equiv \{\alpha ,\beta \}$ is the $%
4^{th}-dimensional$ Fourier transform of $\chi _{S}\{\eta ,\xi \}$. By a
lengthy application of operator algebra and integral calculus we could
evaluate analytically in closed form either $\chi _{_{S}}\{\eta ,\xi \}$\
and $W\{\alpha ,\beta \}$: 
\begin{eqnarray}
W\left\{ \alpha ,\beta \right\} &=&\overline{W}\left\{ \alpha \right\} {\it %
\ }\overline{W}\left\{ \beta \right\} \left[ 1+\left| e^{i\Phi }\Delta
\left\{ \alpha \right\} +\Delta \left\{ \beta \right\} \right| ^{2}-\right. 
\nonumber \\
&&-\left. (\left| \gamma _{A+}\right| ^{2}+\left| \gamma _{A-}\right|
^{2}+\left| \gamma _{B+}\right| ^{2}+\left| \gamma _{B-}\right| ^{2})\right]
\end{eqnarray}
where $\Delta \{\alpha \}\equiv {\frac{1}{2}}[|\gamma _{A+}|^{2}-|\gamma
_{A-}|^{2}-i%
%TCIMACRO{\func{Re}}%
%BeginExpansion
\mathop{\rm Re}%
%EndExpansion
(\gamma _{A+}\gamma _{A-}^{\ast })]$ is given in terms of the squeezed
variables: $\gamma _{A+}\equiv (\alpha _{1}+\alpha _{2}^{\ast })e^{-g}$; $%
\gamma _{A-}\equiv i(\alpha _{1}-\alpha _{2}^{\ast })e^{+g}$. Analogous
expressions involving $B$ and $\beta $ are given by the substitutions: $%
A\rightarrow B$, $\alpha \rightarrow \beta $. The Wigner functions $%
\overline{W}\{\alpha \}\equiv \pi ^{-2}\exp (-[|\gamma _{A+}|^{2}+|\gamma
_{A-}|^{2}])$; $\overline{W}\{\beta \}\equiv \pi ^{-2}\exp (-[|\gamma
_{B+}|^{2}+|\gamma _{B-}|^{2}])$, definite positive over the $4-dimensional$
spaces $\{\alpha \}$ and $\{\beta \}$, represent the effect of the {\it %
squeezed-vacuum}, i.e. emitted respectively by OPA$_{A}$\ and OPA$_{B}$ in
absence of any injection. Inspection of Eq. 3 shows that precisely the
superposition character implied by the {\it entangled} nature of the
injected state $|\Phi \rangle $, Eq. 1, determines through the modulus
square term the $\Phi -dependent$ dynamical quantum interference of the
devices $OPA_{A}$ and $OPA_{B}$, the ones that otherwise act as {\it %
uncoupled} and{\it \ mutually independent, ''}macroscopic{\it '' }objects.

Turn now the attention to NEIF, i.e. to the {\it parity-selective}
interferometric part of our system which operates over the output beams $%
{\bf k}_{j}$ emerging from the QIOPA\ amplifier \cite{11}. Note first that
within the present work NEIF\ reproduces exactly the Bell-state measurement
configuration at the Alice's site of our original quantum state
teleportation (QST)\ experiment \cite{12}. Consider the field emitted by the
NL crystal after R-amplification: Fig. 1. The two beams associated with
modes ${\bf k}_{j}$ ($j=1,2$) are generally phase shifted $\Delta _{j}=(\psi
_{jh}-\psi _{jv})$ by two equal birefringent plates $\Delta _{j}$ and the $%
\pi -$polarizations are rotated by\ two equal Fresnel-Rhomb $\pi -$rotators $%
R_{j}(\theta )$ by angles $\theta _{j}$ respect to directions taken at $%
45^{0}$ with the horizontal. The beams are then linearly superimposed by a
beam splitter (BS) and coupled by two polarizing beam splitters (PBS) to
equal EGG\ SPCM-AQR14 Si-avalanche detectors $D_{1h}$, $D_{1v}$, $D_{2h}$, $%
D_{2v}$ which measures the $(h)$ and $(v)$ $\pi -$polarizations on the
output single modes associated with the field $\widehat{d}_{1h}$, $\widehat{d%
}_{1v}$, $\widehat{d}_{2h}$, $\widehat{d}_{2v}$. A computer controlled mount
allows micrometric displacements of BS along the axis {\bf X}. Consider the
rate of {\it double coincidences:} $(D_{1h}D_{2v})\equiv \langle \Phi |%
\widehat{{\it N}}_{1h}\widehat{{\it N}}_{2v}|\Phi \rangle $ = $%
(D_{2h}D_{1v}) $, where :$\widehat{{\it N}}_{1h}\equiv $ $\widehat{d}%
_{1h}^{\dagger }\widehat{d}_{1h}$. By a detailed account of the full set of
transformations induced by the overall system on the input state $|\Phi
\rangle $ we get: 
\begin{eqnarray}
(D_{1h}D_{2v}) &=&\frac{1}{4}[1+\cos (\Delta -\Phi )]\sin ^{2}(\theta
_{1}+\theta _{2})+S^{2}\times  \nonumber \\
&&\{1+\cos \Phi +\frac{1}{4}[\cos (2\theta _{1})+\cos (2\theta _{2})]^{2}+ 
\nonumber \\
&&-\frac{1}{4}[\cos ^{2}(2\theta _{1})+\cos ^{2}(2\theta _{2})]\cos \Phi +%
\frac{1}{2}\sin ^{2}(\theta _{1}+\theta _{2})\times  \nonumber \\
&&[5+3\cos \Delta +\cos (\Delta -\Phi )-\cos \Phi ]\}+O(S^{4})
\end{eqnarray}
with: $\Delta \equiv (\Delta _{1}-\Delta _{2})$. We may check that the {\it %
phase} $\Phi $ of the input state indeed critically determines the value of
this quantity, e.g. by setting $\Delta =0$, $(\theta _{1}+\theta _{2})={%
\frac{1}{2}}\pi $ the rate reaches its maximum value $(D_{1h}D_{2v})\simeq {%
\frac{1}{2}}$ for any input {\it even-parity} state, e.g. a {\it triplet,} $%
\Phi =0$ while is zero for any input {\it odd-parity }state, $\Phi =\pi $.
This is shown by the data given in Fig. 2 as function of the position {\bf X}
of the BS. %\begin{figure}[tbp]
%\epsfxsize=3.4in \epsfysize=2.6in
%\epsffile{PRL-ParitySelec_Fig_2.eps} \vglue 0.2cm \caption{Parity
%selectivity by the double coincidence $(D_{1h}D_{2v})$.}
%\label{PRL-ParitySelec_Fig_2.eps}
%\end{figure}
There the width of the resonance expresses the coherence time $(225fs)$ of
the detected photons which is determined in the experiment by the passband $%
(\Delta \lambda =2nm)$ of the equal gaussian IF filters placed in front of
the $D^{\prime }s$. Of course the maximum parity-selectivity is realized
when the value of ${\bf X}${\bf \ (}${\bf X\approx 0}$, in Fig. 2{\bf ) }%
realizes the {\it in} {\it principle} indistinguishability of the Feynman
paths affecting the dynamics of each correlated photon couple before
detection. The detailed analysis shows that by adoption of the complementary
coincidences $(D_{1h}D_{1v})$ = $(D_{2h}D_{2v})$ the phase-selectivity
properties given by Eq. 4 are {\it inverted}, viz. there an input {\it %
singlet}\ leads to a resonance {\it peak }in Fig.2{\it ,} a{\it \ triplet }%
to a{\it \ dip} etc. \cite{12}. Changes of the selectivity properties can be
also realized by appropriate settings of $\Delta _{i}$ and $\theta _{i}$
according to Eq. 4.

An important and unexpectedly large $1^{st}-order$ quantum interference
phenomenon was found when the position {\bf Z} of the mirror $M_{UV}$ was
adjusted to realize the {\it time superposition} of the back reflected UV
pulse wave-packet (wp) with wl $\lambda _{p}$ and of the back-reflected\
SPDC generated wp's with wl $\lambda $. A sinusoidal interference fringe
pattern with periodicity $=\lambda $ and {\it visibility} $V$ up to 40\%,
was revealed by the $D^{\prime }s$ within either single detector and
multiple coincidence measurements: cfr. inset of Fig.2. We explain this
striking effect as the realization of the {\it in principle}
indistinguishability, for any detector's frame, of the two possible
directions over which the detected entangled photon couple was originally
emitted: the couple could have been R-generated by the back reflected UV
pulse or L-generated by SPDC and then back reflected \cite{14}. 
%\begin{figure}[tbp]
%\epsfxsize=3.4in \epsfysize=2.6in
%\epsffile{PRL-ParitySelec_Fig_3.eps} \vglue 0.2cm
%\caption{Coincidence rate showing the amplification of an injected
%2-photon entangled singlet state as function of the superposition
%time of the injected and pump wavepackets.}
%\end{figure}
Apart from its relevance and novelty, because of its first realization with
an {\it entangled state}, this effect was helpful to determine the\ value of 
${\bf Z}${\bf \ }corresponding to the {\it maximum} R-amplification: ${\bf Z}%
\approx 0$ in\ both Figs. 2 and 3. The main R-amplification was carried out
with $\Delta =0$, $(\theta _{1}+\theta _{2})={\frac{1}{2}}\pi $ and
investigated by a measurement configuration close to the one $(D_{2v}D_{2h})$
just considered. Precisely, it was found convenient to adopt a related more
complex scheme expressed by the concidence rate: $\Re ({\bf Z})$= $%
(D_{2v}^{\prime }D_{2v}^{"}D_{2h})XOR(D_{1h}+D_{1v})$. This one consists of:
(a) A {\it triple} {\it coincidence} involving 2 detectors $D_{2v}^{\prime }$%
, $D_{2v}^{"}$coupled to the output field $\widehat{d}_{2v}$ by a normal
50/50 BS: Fig.1 inset. (b) This triple coincidence was taken in {\it %
anti-coincidence} with either $D_{1h}$ or $D_{1v}$. These options are
justified as follows: (a) In Eq. 4 the amplified contribution $\varpropto
S^{2}$ had to be discriminated against the dominant first term arising from
the input {\it non amplified} single photon couples: this is obtained by the
BS technique as shown by \cite{15}. (b) The ''{\it noise''} coincidence rate
due to the output {\it squeezed vacuum }is found: $(D_{2v}^{\prime
}D_{2v}^{"}D_{2h})_{vac}$=${\frac{1}{2}}S^{2}(1-\cos \Delta )\cos
^{2}(\theta _{1}-\theta _{2})+$ $S^{4}$. Since the last term arises from 
{\it double} detections by $D-$couples involving either $D_{1h}$ and $D_{1v}$%
, its effect was eliminated by the XOR operation. All this leads, for $%
\Delta =0$ to a theoretical noise output value $(\Re )_{vac}$= 0, viz. {\it %
virtually} to $S/N=$ $\infty $. Of course this condition implies an {\it %
ideally perfect} alignment of the NEIF, i.e. leading to a 100\% visibility ($%
V$)\ of the patterns shown in Fig. 2. In practice the value $V\simeq 0.95$
could be attained so far.

The peak of the signal $\Re ({\bf Z})$ at ${\bf Z}\approx 0$ reported in
Fig.3 shows the evidence of the QIOPA amplification of the {\it %
quantum-injected,}\ $\pi -${\it entangled} state $\left| \Phi \right\rangle
_{s}$ into the\ generation of the {\it multiparticle}, $\pi -${\it entangled}
Schroedinger Cat state: $\left| \Phi \right\rangle _{OUT}=2^{-{\frac{1}{2}}%
}[|\Psi _{A}\rangle -|\Psi _{B}\rangle ]$. Of course the beast is presently
small because of the small value of the gain, evaluated on the basis of the
properties of the NL\ crystal: $g\approx 0.22$. By a series of comparative
measurements, carried out in presence and absence of the input UV-{\it pump (%
}with wl $\lambda _{p}$) and {\it ebit-injection} beams (wl $\lambda $), it
was found that each input quantum injected photon couple {\it QED} {\it %
stimulated }an{\it \ }average{\it \ }number of additional couples $N=0.20$.
With our present experimental conditions the system generates odd-parity
entangled $4-photon$ states $\left| \Phi _{2}\right\rangle =$ $2^{-\frac{1}{2%
}}[\left| 22.00\right\rangle -\left| 00.22\right\rangle ]\equiv $ $2^{-\frac{%
1}{2}}[\left| \uparrow \uparrow \right\rangle _{1}\left| \downarrow
\downarrow \right\rangle _{2}-\left| \downarrow \downarrow \right\rangle
_{1}\left| \uparrow \uparrow \right\rangle _{2}]$ at a rate $\approx 3\times
10^{3}\sec ^{-1}$. This results can be {\it linearly scaled} however e.g. by
adoption of a more efficient NL crystal and of a more powerful UV source. In
the near future at least a factor $17$ increase of the value of\ $g$ shall
be attained by\ the adoption within our system of a standard Ti:Sa
Regenerative Amplifier Coherent REGA9000 operating with $\delta
_{t}=150f\sec $ pulses, a 270 ${\it Khz}$ rep-rate and an average UV\ output
power $\approx $0.30 W: an apparatus already installed in our laboratory. In
this case the gain parameter will increase up to a value very close to its
maximum: $\Gamma \equiv \tanh g$ $\approx 1$ and the average number of
photon couples {\it QED stimulated }by any single injected ebit will be very
large $N\gg 1$, as implied by the explicit expressions of the entangled
''macrostates'' $|\Psi _{A}\rangle $ and $|\Psi _{B}\rangle $ given in Eq.
2. In conclusion we have reported the succesful implementation of an {\it %
entangled} Schroedinger Cat apparatus. If the present results will prove to
be scalable as expected, the present realization would open a new era of
basic investigations on the persistence of the validity of several crucial
laws of quantum mechanics for systems of increasing complexity, in a
virtually {\it decoherence-free} environment \cite{8,9}. The first test we
are performing with the present system concerns the violation of Bell-type
inequalities in the multiparticle regime: a long sought new perspective in
the fundamental endeavour on quantum nonlocality \cite{16}. We acknowledge
enlightening discussions and collaboration with S.Branca, V.Mussi, F.Bovino,
M.Lucamarini. We thank MURST and INFM (Contract PRA97-cat) for funding.

\begin{figure}[tbp]
\caption{Experimental apparatus.}
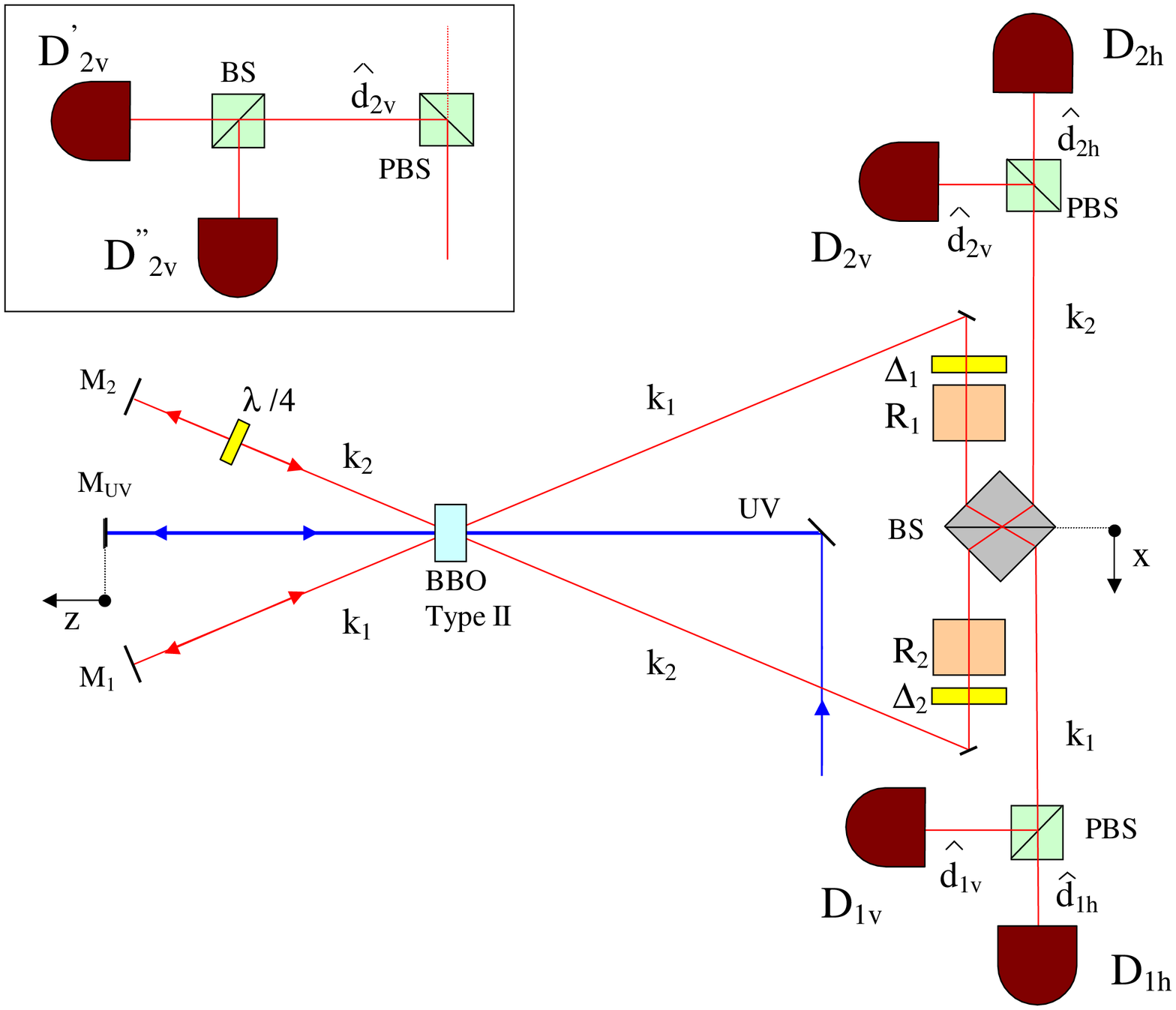
\label{PRL-ParitySelec_Fig_1.eps}
\end{figure}

\begin{figure}[tbp]
\caption{Parity selectivity by the double coincidence $(D_{1h}D_{2v})$.}
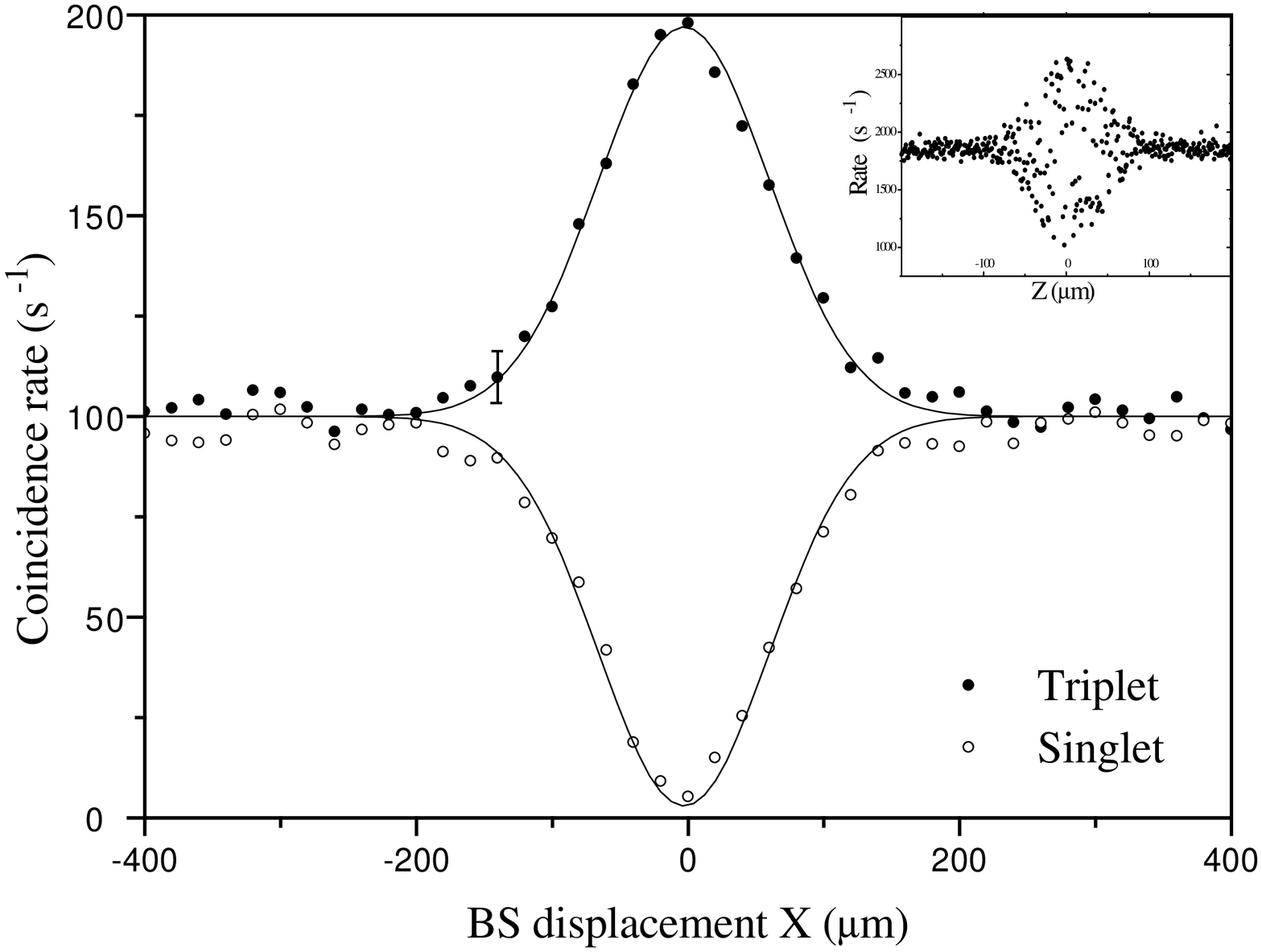
\label{PRL-ParitySelec_Fig_2.eps}
\end{figure}

\begin{figure}[tbp]
\caption{Coincidence rate $\Re$ showing the amplification of an injected
2-photon entangled singlet state as function of the superposition time of
the injected and pump wavepackets.}
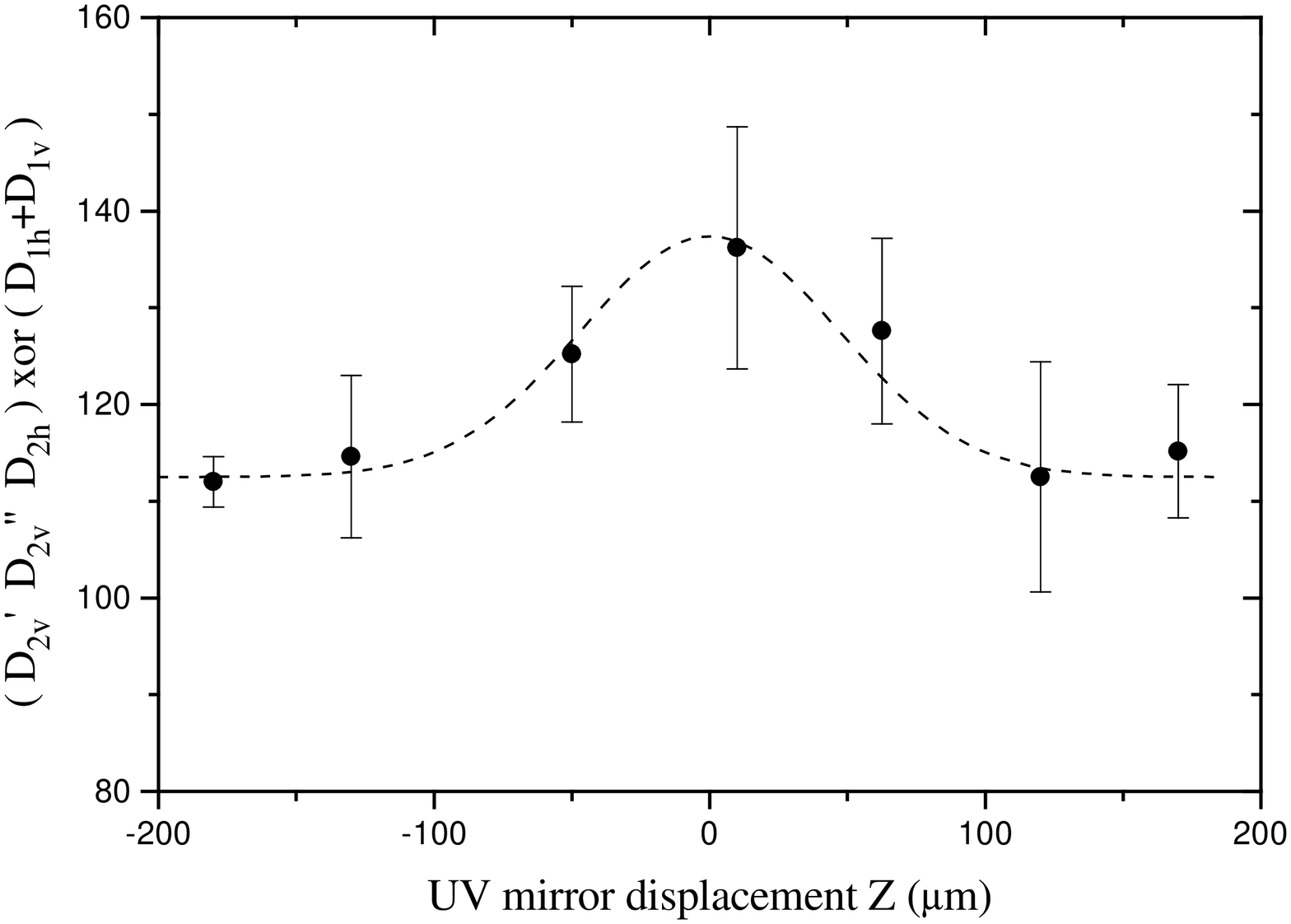
\label{PRL-ParitySelec_Fig_3.eps}
\end{figure}

\end{document}